# Laser powder-bed fusion additive manufacturing: physics of complex melt flow and formation mechanisms of pores, spatter, and denudation zones

Saad A. Khairallah, Andrew T. Anderson, Alexander Rubenchik, Wayne E. King

Lawrence Livermore National Laboratory, 7000 East Ave. Livermore, CA 94550

Corresponding author: Saad A. Khairallah. Tel: 1-925-422-0675.

E-mail addresses: khairallah1@llnl.gov (Saad A. Khairallah), anderson1@lln.gov (Andrew Anderson), rubenchik1@llnl.gov (Alexander Rubenchik), king17@llnl.gov (Wayne E. King)

**ABSTRACT**

This study demonstrates the significant effect of the recoil pressure and Marangoni convection in laser powder bed fusion (L-PBF) of 316L stainless steel. A three-dimensional high fidelity powder-scale model reveals how the strong dynamical melt flow generates pore defects, material spattering (sparking), and denudation zones. The melt track is divided into three sections: a topological depression, a transition and a tail region, each being the location of specific physical effects. The inclusion of laser ray-tracing energy deposition in the powder-scale model improves over traditional volumetric energy deposition. It enables partial particle melting, which impacts pore defects in the denudation zone. Different pore formation mechanisms are observed at the edge of a scan track, at the melt pool bottom (during collapse of the pool depression), and at the end of the melt track (during laser power ramp down). Remedies to these undesirable pores are discussed. The results are validated against the experiments and the sensitivity to laser absorptivity is discussed.

**1. Introduction**

Additive manufacturing (AM) is paving the way toward the next industrial revolution [1]. The essence of this advancement is a part that is produced from a digital model by depositing material layer by layer, in other words, 3D printing the model. This technique is in contrast with the traditional subtractive and formative manufacturing approaches. It also eliminates most of the constraints that hinder optimal design, creativity and ease of manufacturing of complex parts [2] [3].

A promising future is in store for L-PBF AM. However, widespread adoption of L-PBF with metallic parts hinges on solving a main challenge: the requirement that the final product should meet engineering quality standards [4]. This includes reducing porosity, since pore defects have one of the most adverse effect on mechanical properties. Experimental advances on this front rely on trial and error methods, which are costly and time inefficient. An attractive alternative to answering this challenge is through modeling and predictive simulation.

The finite element method (FEM) is the most popular numerical method for simulation of metal powder bed additive manufacturing processes. Critical reviews by Schoinochoritis *et al* [5] and King *et al.* [6] discuss different FEM models, assumptions and results. The emphasis is how to get the most out of FEM simulations while avoiding computational expense. Some simplifications include (1) treating the powder as a homogeneous continuum body with effective thermomechanical properties (2) treating the laser heat source as a homogeneous model that deposits laser energy volumetrically like with De-Beer-Lambert's law or one derived for deep powder bed [7], and (3) ignoring melt pool dynamics and therefore assuming a steady state. Take for example the work of Gu *et al.* [8] who employ a commercial code based on the finite volume method (FVM) to highlight the significant effect of Marangoni convection on heat and mass transfer in a continuum 3D model. In that model, the discrete nature of the powder is not accounted for; hence the melt flow is symmetric along the melt track and does not exhibit fluctuations that may be introduced by a randomly packed powder bed.

The current paper falls outside the FEM body of work. Our approach is to study the L-PBF problem with a fine-scale model that treats the powder bed as randomly distributed particles. There are few studies that follow this mesoscopic approach.

In [9], Gutler et al. employ a volume of fluid method (VOF) and were the first to show more realism with a 3D mesoscopic model of melting and solidification. However, a single size powder arranged uniformly was represented at a coarse resolution that does not resolve the point contacts between the particles. The paper makes qualitative correlations with experiments.

Körner et al. [10] use the lattice Boltzmann method (LBM) under the assumption that the electron beam melting process can be represented in 2D. One big hurdle in this method is the severe numerical instabilities occurring when accounting for the temperature. Körner uses the multi-distribution function approach to reduce these limitations under the assumption that the fluid density is not strongly dependent on temperature. The method has been applied in 2D to study single layer [11] and layer upon layer consolidation [12], and shows the importance that the powder packing has on the melt characteristics. Their observation of the undesirable balling effect was attributed to the local powder arrangement [11]. Recently, a 2D vapor recoil pressure model was added in [13] to improve the melt depth predictions. The Marangoni effect is neglected. In [14], a 3D model that does not include recoil, Marangoni, or evaporation effects was used to establish process strategies suitable to reduce build time and cost while enabling high-power electron beam applications.

Khairallah *et al.* in [15] reported on a highly resolved model in 3D that considers a powder bed of 316L stainless steel with a size distribution taken from experimental measurements. Khairallah *et al.* emphasized the importance of resolving the particle point contacts to capture the correct reduced effective thermal conductivity of the powder and the role of surface tension in breaking up the melt track into undesirable ball defects at higher laser scan speeds due to a variant of Plateau-Rayleigh instability theory [16].

A recent mesoscopic study by Lee and Zhang [17] introduces the powder into the model using the discrete element method. Their VOF study emphasizes the importance of particle size distribution and discusses the smoothing effect of small particles on the melt. They agree with Khairallah *et al.* [15] that balling is a manifestation of Plateau-Rayleigh instability and add that higher packing density can decrease the effect. Recoil and evaporation effects are neglected.

Recently, Qiu *et al.* [18] performed an experimental parameter study, whereby the surface roughness and area fraction of porosity were measured as a function of laser scan speed. They noted that the unstable melt flow, especially at high laser scan speed, increases porosity and surface defects. Based on a CFD study of regularly packed powder of a single large size of 50 μm, they believe that the Marangoni and recoil forces are among the main driving forces for the instability of melt flow.

This manuscript describes a new high fidelity mesoscopic simulation capability developed to study the physical mechanisms of AM processes by eliminating certain physical assumptions that are prevalent in the literature due to modeling expense. The model uses a laser ray tracing energy source and is in 3D to account for the fluid flow effects due to the recoil pressure, the Marangoni effect, and evaporative and radiative surface cooling. The new findings point out the importance of the recoil pressure physics under the laser and its dominant effect on creating a topological depression (similar to a keyhole) with complex strong hydrodynamic fluid flow coupled to a Marangoni surface flow. A vortex flow results in a cooling effect over the depression, which coupled to evaporative and radiation cooling over an expanded recoiled surface, regulates the peak surface temperatures. This finding should benefit part scale and reduced order modeling efforts, among others, that limit heat transfer to just conduction and therefore suffer from uncontrolled peak surface temperatures and may have to resort to model calibration to capture the effect.

This study, other than detailing the dominant physics in L-PBF, reveals the formation mechanisms for pore defects, spatter, and the so-called denudation zone where powder particles are cleared in the vicinity of the laser track. Several authors report experimentally observing these effects, however, they formulate assumptions for formation mechanisms since, experimentally, it is challenging to dynamically monitor the L-PBF process at the microsecond and micrometer scales. For example, Thijs et al. assume that some particles located in the denudation zone melt incompletely and create pore defects [19] and that other pores form due to the collapse of a keyhole [20]. Qiu et al. [21] observe open pores and assume that the incomplete re-melting of the previous layer generates spherical pores.

The present study explains how three kinds of pore defects (depression collapse, lateral pores, open and trapped pores) are generated and discusses strategies to avoid them. This study, thanks to the laser ray tracing energy source and the inclusion of recoil pressure, is also able to describe the physical mechanisms behind sparking [22], spattering, and denudation [23] [24].

Experimental validation with sensitivity to the choice of laser absorptivity is also presented. The model makes use of the ALE3D [25] massively-parallel multi-physics code. Code and material property details can be found in [15] [26].

## 2. Model: Underlying Physics and Validation

*2.1 Volumetric versus ray tracing laser heat source*

L-PBF is a heat driven process, which needs to be modeled accurately. This study uses a ray tracing laser source that consists of vertical rays with a Gaussian energy distribution (D4σ = 54 μm). The laser energy is deposited at the points of powder-ray intersections. To reduce the computational complexity, the rays are not followed upon reflection. The direct laser deposition is an improvement over volumetric energy deposition (energy as a function of fixed Z-axis reference) used commonly in the literature. Firstly, in reality the heat is generated where the laser rays hit the surface of the powder particles and diffuses inward, whereas homogeneous deposition heats the inner volume of the particle uniformly. Secondly, the rays track the surface and can reproduce shadowing. In Figure 1a, a 150W Gaussian laser beam is initially centered above a 27 μm particle sitting on a substrate and moved to the right at 1 m/s. For volumetric energy deposition, melting happens simultaneously everywhere inside the particle. The wetting contact with the substrate increases rapidly, which artificially increases heat dissipation. On the other hand, with realistic laser ray tracing, melting is non-uniform as it occurs first at the powder particle surface. More heat accumulates inside the powder particles compared with the homogeneous laser deposition because it releases to the substrate slowly through a narrow point contact. If insufficient heat is deposited, the particles are partially melted and contribute to surface and pore defects as discussed in section 3.2.5. The laser ray tracing heat source helps to better couple the physics behind surface heat delivery and melt hydrodynamics.

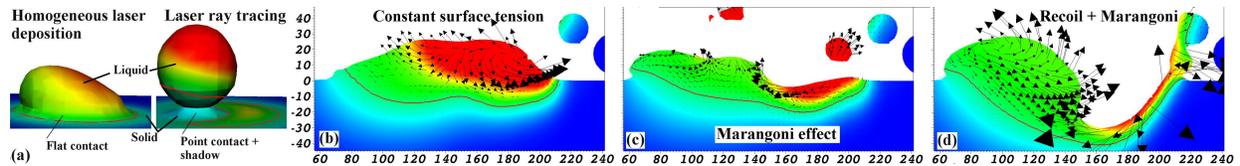

**Figure 1. Incremental physics fidelity, significantly alters the heat transfer, melt pool depth and flow. The red pseudocolor corresponds to temperature scale capped at 4000K, blue is 293K. The red contour line is the melt line. The powder particle is illuminated by a laser moving to the right for 10 μs. The melt tracks are 2D slices of 3D simulations demonstrating the effect of improved physics modeling on the melt pool (see section 2.2 and 3).**

*2.2 Temperature driven 3D flow effects: Surface tension, Marangoni convection, and recoil pressure*

Figure 1b, c and d illustrate the significant change of melt pool characteristics as more temperature dependent physics is included. If surface tension is assumed to be temperature independent, unphysical effects are observed. The melt pool is the shallowest with a constant surface tension in Figure 1b and shows a balling effect due to surface tension tendency to minimize surfaces by creating liquid spheres. The melt flow is also driven by buoyancy.

Next in Figure 1c, the strong temperature gradients below the laser necessitate enabling temperature dependent surface tension, which creates Marangoni effects. It drives the melt flow from the hot laser spot toward the cold rear. This serves to increase the melt depth, recirculate the melt flow (hence cool the location of the laser spot) and create spattering as liquid metal with low viscosity ejects away from the surface.

The next increment in physics fidelity in Figure 1d comes from recognizing that the surface temperatures below the laser spot can easily reach boiling values. The vapor recoil pressure adds extra forces to the surface of the liquid that create a melt pool surface depression below the laser. Since the applied heating in L-PBF does not cause extreme vaporization (ablation), the model does not resolve the vapor flow discontinuities and expansion from the liquid phase to ambient gas **[27] [28]**, nor does it include the mass lost to vaporization. In this study, a simplified model due to Anisimov [29] is employed, which has been used previously[26] [30] [31]. The recoil pressure $P$ depends exponentially on temperature, $P(T) = 0.54 P_a exp^{-\frac{\lambda}{K_B}\left(\frac{1}{T}-\frac{1}{T_b}\right)}$, where $P_a$=1 bar is the ambient pressure, $\lambda$=4.3 ev/atom is the evaporation energy per particle, $K_B$=8.617 ×10$^{-5}$ ev/K is Boltzmann constant, $T$ is the surface temperature and $T_b$=3086 K is the boiling temperature of 316L stainless steel. By combining the Marangoni effect with recoil pressure, the melt depth significantly increases, which also increases the surface area of the melt pool (by creating a depression; see section 3.1) and helps further with cooling due to additional evaporative and radiative surface cooling. In fact, among the three 2D melt pool slices, this last figure shows the least amount of stored heat (shown in red pseudocolor).

*2.3 Surface cooling: Evaporative and radiative cooling*

Since it is essential to calculate the surface temperature accurately, extra care is given to account for thermal losses. An evaporative cooling term is calculated at the surface interface and has the big role of limiting the maximum surface temperature under the laser, since the flux of evaporated metal vapor increases exponentially with $T$. According to Anisimov's theory [29], around 18 % of the vapor condenses back to the surface due to large scattering angle collisions in the vicinity of the liquid and hence reduces the cooling effect. The net material evaporation flux is $J_v = 0.82 A P(T)/\sqrt{2\pi M R T}$ and is consistent with the recoil pressure *P(T)* derivation. Here, *A* is a sticking coefficient, which is close to unity for metals, *M* is the molar mass, *R* the gas constant and *T* the surface temperature. The model neglects evaporative mass loss, since the amount is negligible. As a conservative mass loss estimate, consider an area of 1 mm x 54 μm fixed at 3000 K for 0.67 ms. The mass loss amounts to ~0.1 μg, which is much less than the mass of an average stainless steel particle with a radius of 27 μm.

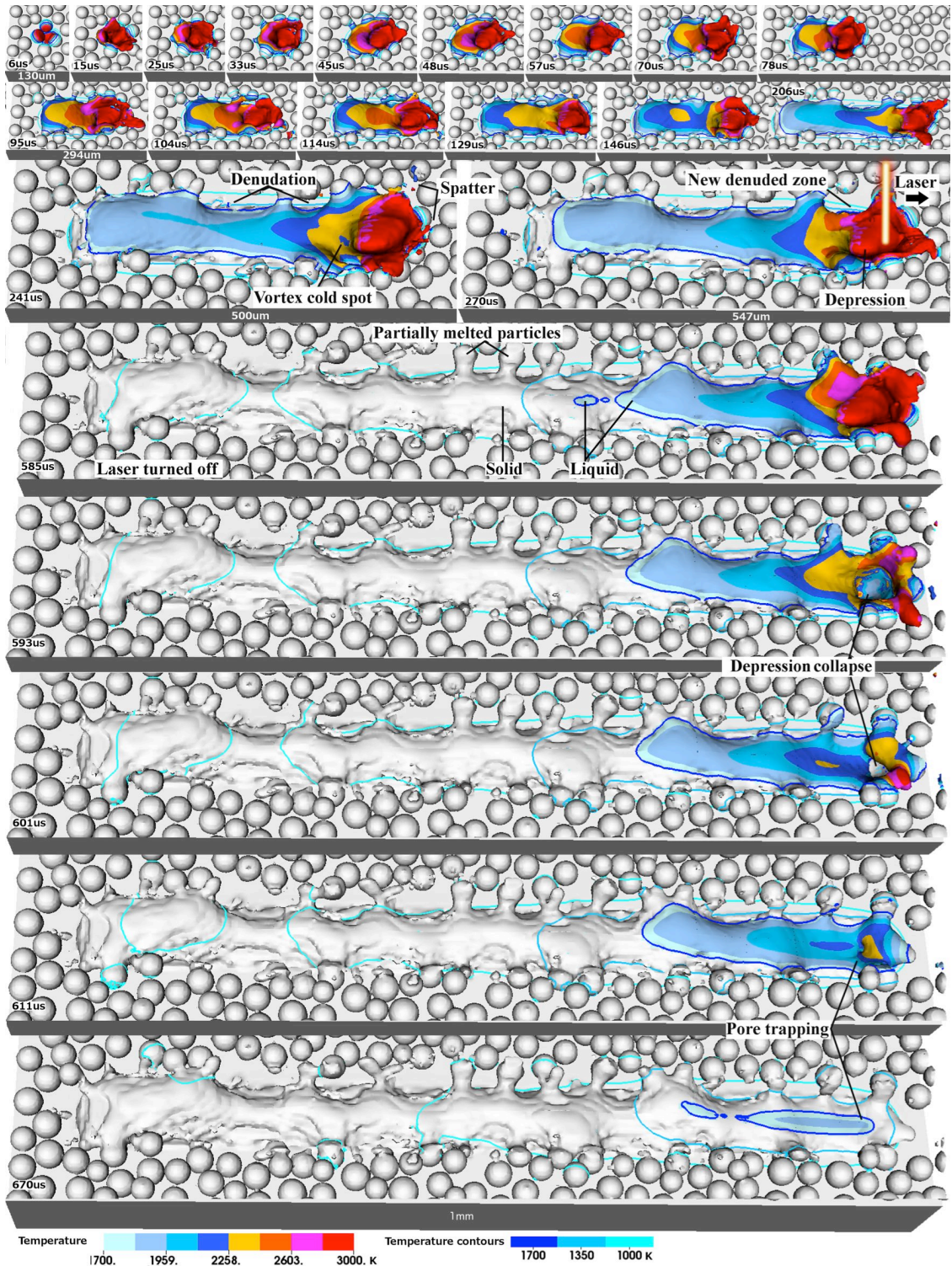

**Figure 2.** Time snapshots showing the evolution of the surface temperature. The laser scan speed is 1.5 m/s and moving to the right with a power of 200 W. The liquid melt pool is confined within the colored regions (T>1700 K). The surface melt reaches a steady state late in time around 229 μs. The laser creates a topological depression, which is the site of forward and sideways spatter, and also contributes to the denudation process. The laser is turned off at 585 μs. Later in time, the depression collapse creates a trapped pore beneath the surface.

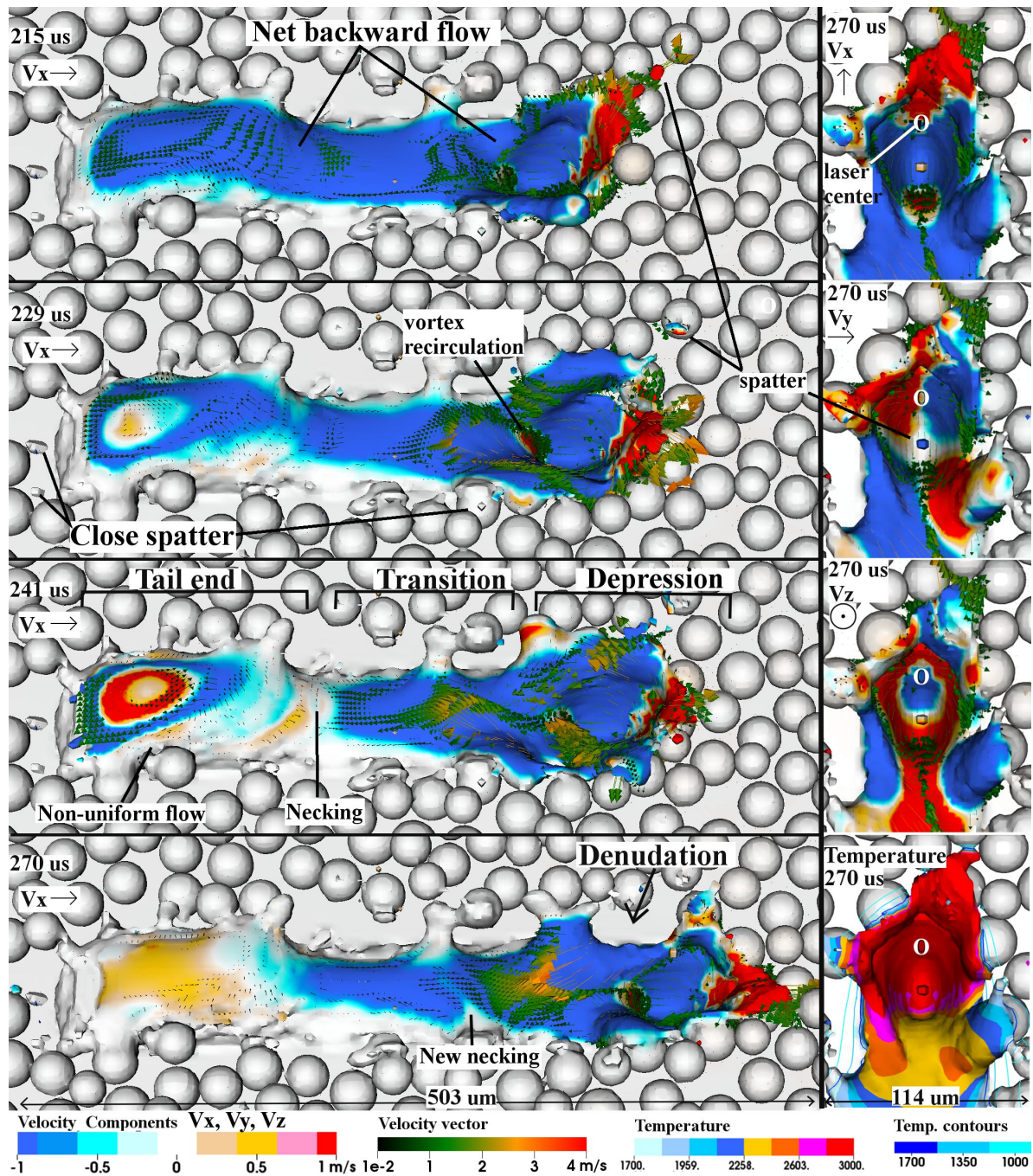

**Figure 3. Time snapshots of the melt flow showing spattering and denudation.** The melt has a large backward flow (blue color; Vx<0) due to Marangoni effect and recoil, compared to forward flow (Vx> 0; red color). The backward net flow breaks up later in time at the necking. The velocity scale is capped at +-1m/s for better visualization. The right panel magnified view at 270μs (flow rotated by +90°) shows the velocity components (Vx, Vy, Vz) and the temperature (with contour lines) at the depression. The white letter O shows that the laser center is not at the bottom of the depression.

In addition to evaporative cooling, radiative cooling that follows the Stefan-Boltzmann law, $\boldsymbol{R = \sigma\varepsilon(T^4 - T_o^4)}$, assuming black body radiation, is included. Note that compared to the total deposited laser energy, the radiation heat losses are quite small. Here the Stephan's constant is σ=5.669 ×10$^{-8}$ W/m$^2$K$^4$. The emissivity, ε, varies with temperature and surface chemistry and therefore is hard to represent **[32]**. For simplicity, an average value for the emissivity is taken to be 0.4 for the solid stainless steel and 0.1 for the liquid state. $T_o$ is the ambient temperature. The model assumes that the lateral sides of the problem domain are insulated, while the bottom surface uses a boundary condition that approximates the response of a semi-infinite slab.

*2.4 Experimental model validation and sensitivity to material absorptivity*

The highest temperature gradients exist soon after the laser is turned on. For a laser power of 200 W and laser scan speed of 1.5 m/s, the surface melt pool shape settles into quasi-steady-state about ~225 μs after the laser is turned on, (See Figure 2). The width of the melt pool is observed to fluctuate along the solidified track. On the other hand, the melt depth increases until it stabilizes earlier at ~100 μs.

| | | | | | |
|---|---|---|---|---|---|
| *P300S1800* *abs. 0.35* | *D68/65* *W96±8/94* | *P200S1200* *abs. 0.35* | *D70/68* *W94±12/104* | *P150S800* *abs. 0.35* | *D69/67* *W89±4/109* |
| *P200S1500* *abs. 0.3* | *D45/57* *W80±5/84* | *P200S1500* *abs. 0.35* | *D54/57* *W80±9/84* | *P200S1500* *abs. 0.4* | *D60/57* *W80±4/84* |

**Table 1.** Simulation and experiment data (separated by /) comparison of depth $D$[µm] and width $W$[µm] at different laser scan speeds $S$[mm/s] and powers $P$[Watts]. The material absorptivity *abs*. is held constant in the first row. The experimental uncertainty is 5 µm and the simulation's melt depth is on the order of zone size, which is 3 µm. The width fluctuates more than depth. The second row tests the sensitivity of the results to the absorptivity. An absorptivity of 0.35 shows the best agreement with the experiment.

Table 1 shows that the melt depth, for a constant absorptivity of 0.35, yields very good quantitative agreement with the experiment. The second row shows a sensitivity study of melt pool depth and width on absorptivity. The melt pool depth is sensitive to laser absorptivity, whereas the width does not vary much as it depends mostly on beam size. Taking a constant absorptivity (which is a common approach [33]), is a main approximation in the model. A depression forms below the laser that could absorb more heat due to multiple reflections (see depressions in Figure 2). Experimentally, a plasma/metal vapor plume can change the absorptivity along the pool depth. However, incorporating a variable absorptivity is quite complex and not necessary for this model since the depression is not as deep as a keyhole [30].

## 3. Results and Discussions

*3.1 Anatomy of a melt track*

It is possible to subdivide the melt track into three differentiable regions: a *depression* region located at the laser spot, a *tail end* region of the melt track located near the end, and a *transition* region in between (see Figure 3 at 241 µs). This choice of subdivision is based on the exponential dominance of the recoil force at the depression and the dominance of surface tension in the cooler transition and tail regions.

The depression may be viewed as a source of fluid. While the flow at the depression is complex, the flow in the transition zone has a net surface velocity component (Vx) in the negative direction (to the rear). The velocity snapshots from 215-270 μs in Figure 3 show a dominant blue region (Vx<0) behind the depression region. At 225 μs, the surface melt pool shape achieves a steady state. The backwards flow starts to break up at the tail end of the track. Later (at 241 μs and 270 μs), it becomes easy to distinguish the three regions: the depression, the transition, and the tail. When placed in the laser reference frame, this flow breakup is reminiscent of the Plateau-Rayleigh instability in a cylindrical fluid jet that breaks into droplets, which has been observed in L-PBF experiments [24] [34]. This is a manifestation of nature's way of minimizing surface energy using surface tension. The melt track achieves a lower surface energy by transitioning from the segmented cylinder [24] observed in the transition region to the segmented hemispherical-like tail-end region [16]. The necking locations where the melt track dips or even disappears correspond to the necking of a narrow cylindrical fluid jet prior to break up into droplets. These dips cool down quickly. It is possible to control the magnitude of the fluctuations in the tail-end regions by adjusting the laser speed for a given power, and hence averts major balling, by controlling the heat content over time in the melt track. Less heat content gives the surface tension less time to completely break the flow [15]. For the current simulation parameters (scan speed lower than in [15]), the balling instability is mild.

*3.2 Effects of a strong dynamical melt flow*

*3.2.1 Depression formation*

Figure 4 shows a time series of track cross sections for a fixed position with the laser moving out of the plane. They highlight the formation of the depression region, which is marked by the highest temperatures achieved on the track (See Figure 2). In this region, which is directly under the laser, the recoil effect is dominant due to its exponential dependence on temperature and creates a noticeable topological depression. At 45 μs, the momentum imparted by hot spatter falling ahead of the depression moves the particles lying ahead of the laser. After 58 μs, the particles melt within 20 μs ahead of the Gaussian laser center. The smaller one melts completely before the larger one and hence increases the particle thermal contact area (see discussion on the laser source in section 2.1 and Figure 1). The ensuing liquid has a large speed lateral flow component ~4-6m/s directed away from the center of the hot spot, which is marked by a narrow black temperature contour line (3500 K). The center of the laser reaches the slice ~30 μs after first signs of powder melting. With surface temperatures approaching the boiling temperature, the recoil pressure applies an exponentially increasing force normal to the surface, which accelerates the liquid away from the center as the velocity vectors show at 76 μs. The result is a depression with a thin liquid boundary layer at the bottom. It is most thin at the bottom of the depression, where the temperature is the highest. The vertical velocity component of the liquid is negative at the bottom of the depression where the recoil force is digging the hole, and is positive along the sidewalls and the rim where the liquid escapes vertically at relatively high speed (~1m/s) and contributes to spattering (as seen in Figure 3 at 270 μs).

This depression is closely related to the keyhole cavity observed in welding [33]. Also, King *et al* [35], experimentally observed keyhole mode melting in laser powder bed fusion and ascribe this to a surface threshold temperature close to boiling. The recoil force is the main driving force for the keyhole mode melting. Many numerical models for keyhole mode laser welding involve simplifying assumptions. They typically balance the recoil force, the surface tension pressure, and hydrostatic liquid pressure. Furthermore, the models can be 2D and often consider heat transfer by conduction only, without accounting for the influence of convection on heat dissipation. Since similar underlying physics processes also occur in L-PBF, these simplifying approaches have also been adopted when developing L-PBF models [5]. However, missing the effects, such as convective cooling, of the strong dynamical flow shown in Figures 3 and 4 may limit the range of predictability of these models.

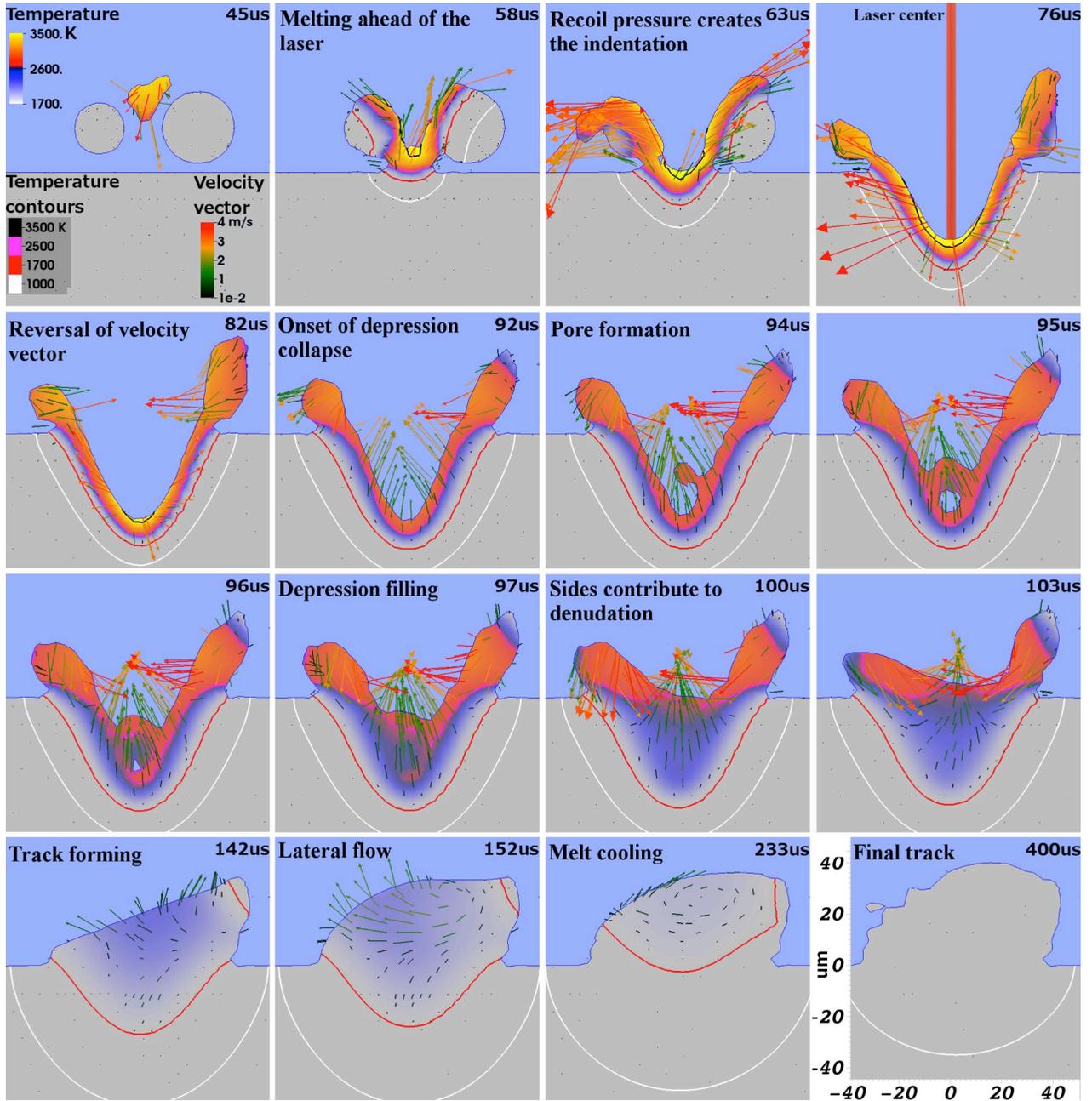

**Figure 4.** Lateral 2D slices showing the temperature and velocity field of the melt as the laser scans (direction out of page) by a fixed location. They show the events before the arrival of the laser center (45-76 μs), the indentation formation (76-82 μs), the indentation collapse and formation of a pore (92-103 μs) and the asymmetrical flow pattern due to an asymmetrical cooling as the melt solidifies (142-400 μs).

*3.2.2 Depression collapse and pore formation mechanism*

At 82-92 μs (Figure 4), the laser's hottest spot has just passed through the plane of the figure. The temperature at the back of the depression decreases, which is indicated by the recession of the black temperature contour line (~3500K). Behind the hottest spot, a decrease in temperature is accompanied by an exponential decrease in recoil force; however, the surface tension increases at lower temperatures and overcomes the recoil force effect, which was keeping the depression open. As a result, the melt-flow velocity-vector field reverses direction towards the center in Figure 4 starting at 82 μs. This reversal is abrupt and causes the sidewall to collapse within 5 μs. Gravity is included in the model but has negligible effect on this timescale. This fast flow increases the chance of trapping gas bubbles and therefore forming pores at the bottom of the track. The sequences at 94-97 μs show this pore formation mechanism.

Figure 5a shows another possible mechanism for pore formation due to a vortex, represented by a velocity vector field circulating counter-clockwise that follows the depression closely from the rear. We speculate that it could trap bubbles and/or seed a bigger pore by pore coalescence meanwhile the solid front advancing from the bottom would catch the bubble and freeze it into a permanent pore.

The vortex has another effect. It helps with cooling as it brings colder liquid back to the depression. The vortex is visible in Figure 3 (270 μs) as a small red patch ($V_x>0$), at the back wall of the depression, surrounded by a blue region (Vx < 0). Figure 5-a) shows cold temperature contour lines pushing hotter ones towards the depression. Figure 2 (241 µs) shows this cooling effect as a cold blue patch (T < 2258K) mixing with hotter yellow region. The vortex only ceases to exist after the laser is turned off at 585 µs.

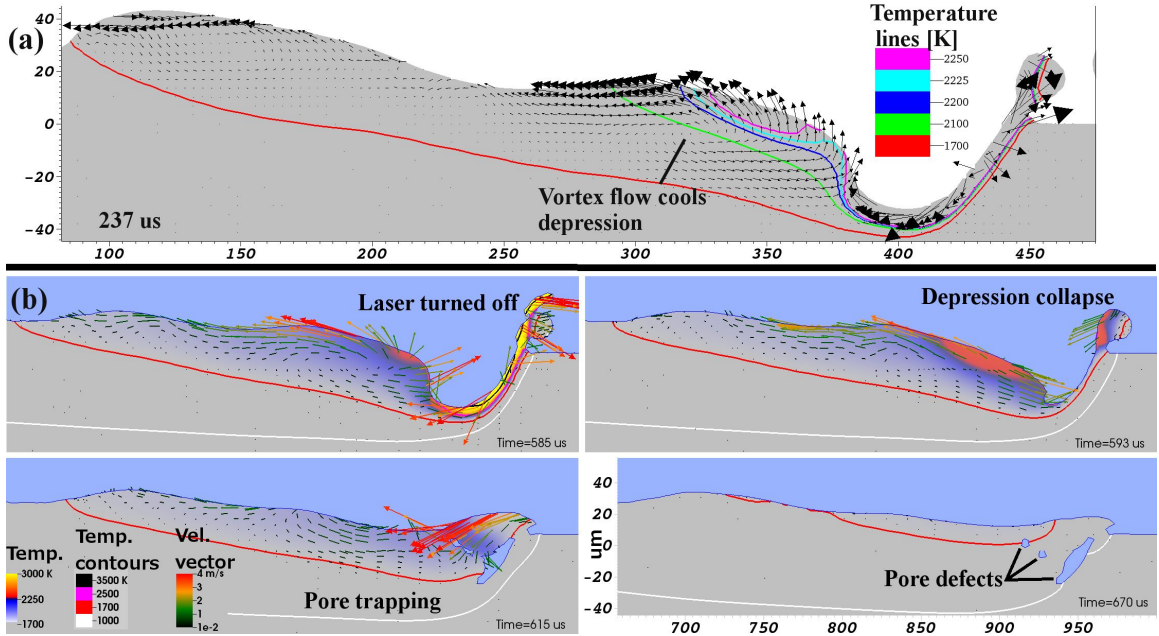

**Figure 5.** Longitudinal 2D slices showing velocity field and temperature. In a), the velocity field (maximum magnitude 9 m/s) shows a vortex pushing cold temperature contour lines into hotter regions behind the depression zone. In b), the figures at 585-670 µs show another process where pores are formed. Upon turning the laser off, the depression collapse creates three pores.

King et al. **[35]** observed pores in keyhole-mode laser melting in laser powder bed fusion experiments on 316L stainless steel. King et al. followed a similar scaling law as Hann et al. **[36]** to analyze their findings. Hann et al. derived a scaling law to classify a variety of materials with different welding process parameters. The general welding data seem to collapse to one curve under the assumption that the melt depth divided by the beam size is a function of $\Delta H/h_s$, which is the deposited energy density divided by the enthalpy at melting. King et al. showed that similar scaling applied well to laser bed fusion and found that the threshold to transition from conduction to keyhole mode laser melting is $\Delta H/h_s \approx (30 \pm 4)$. They concluded that "going too far below the threshold results in insufficient melting and going too far above results in an increase in voids due to keyhole mode melting". With a ratio of $\Delta H/h_s = 33$, the simulation model in this study is at the threshold, and indeed it shows a relatively small keyhole like depression and small number of pores as is evident from the 3D view from below in Figure 6.

*3.2.3 Denudation mechanism*

At 100-400 µs in Figure 4, the liquid fills the depression and grows in height. A lateral liquid flow is noted due to asymmetrical cooling in the transition region. This is due to partially melted particles (see Figure 2 at 585 µs) that remain in touch with the melt track and dissipate heat laterally. The surface tension will then pull surface fluid towards the cold spot (Marangoni effect) and hence bias any lateral circulation. This is undesirable because these can possibly create bridges with gaps underneath and seed further defects in the next deposited layer.

Most often, the side particles melt completely and are trapped in the flow in the transition region. The cause is liquid that circulates around the rim of the depression and resembles a teardrop. This pattern is observed in traditional welding. It is visible in Figure 3 (270 µs, Vy) where the flow alternates between red (Vy<0) and blue (Vy>0) two times around the depression rim: Once ahead of the depression, to indicate motion away from the laser spot, and one last time to indicate fluid coming from the sides and joining to form the transition region.

This circular motion has a wider diameter than the melt track width. This can be seen in Figure 4 at 100 µs where the melt temperature contour line in the substrate does not extend far enough to contain the melt above. The liquid that spills over to the sides catches the neighboring particles and drags them into the transition zone, behind the depression, hence creating what is known as the denudation zone along the sides of the track **[23] [24]**. The velocity vectors in the snapshot series from 241 μs to 270 μs in Figure 3 show the denudation from top view: The flow at 241 μs overlaps with particles that disappear later at 270 μs. The mechanism for the denudation is enhanced by the high velocity circular flow (1- 6 m/s). Yadroitsev et al. **[23] [24]** observed the denudation zones experimentally and attributed it partially to particles in the immediate vicinity of the track as seen in this study.

*3.2.4 Spatter formation mechanism*

Figure 3 and Figure 4 at 45 μs show the build up of liquid that develops ahead of the depression and the laser spot. This build up is similar in nature to the "bow wave" that develops as a boat moves through the water or to the motion of snow rolling over in front of a snowplow. The liquid colored in red in Figure 3 moves up the front wall of the depression and spills over onto the powder particles ahead of the laser beam. This is an important feature as liquid can be pinched off in this process and be deposited as spatter particles in the powder bed.

Figure 6a details how this liquid build-up (or "bow wave") leads to spattering, which is experimentally observed in [22]. The high vapor surface flux (referred to as gas plume in [22]) exerts a pressure force that ejects liquid metal. When the liquid metal elongates, it thins out and breaks up into small droplets due to surface tension tendency to minimize surface energy. Figures 3 and 6a show the elongation is in the radial direction to the laser spot and pointing away from the melt pool.

*3.2.5 Lateral shallow pores and trapped incompletely melted particles*

Another pore formation mechanism takes place in the transition region. The strong high-speed flow along the depression rim that brings in the particles and hence creates the denudation zone also mixes in voids that originally existed between the particles. One realistic effect of the ray tracing laser source is that it allows for partial melting of particles. If a particle does not melt completely and merge with the melt pool, the voids present between the particles may contribute to pore defects. The snapshots in Figure 6b show a partially melted particle below which, a shallow lateral pore on the order of 5 μm is generated. These trapped particles also are defects that increase surface roughness and "deteriorate the wetting behavior of the next layer and act as the origin of continued layer instability" according to [37].

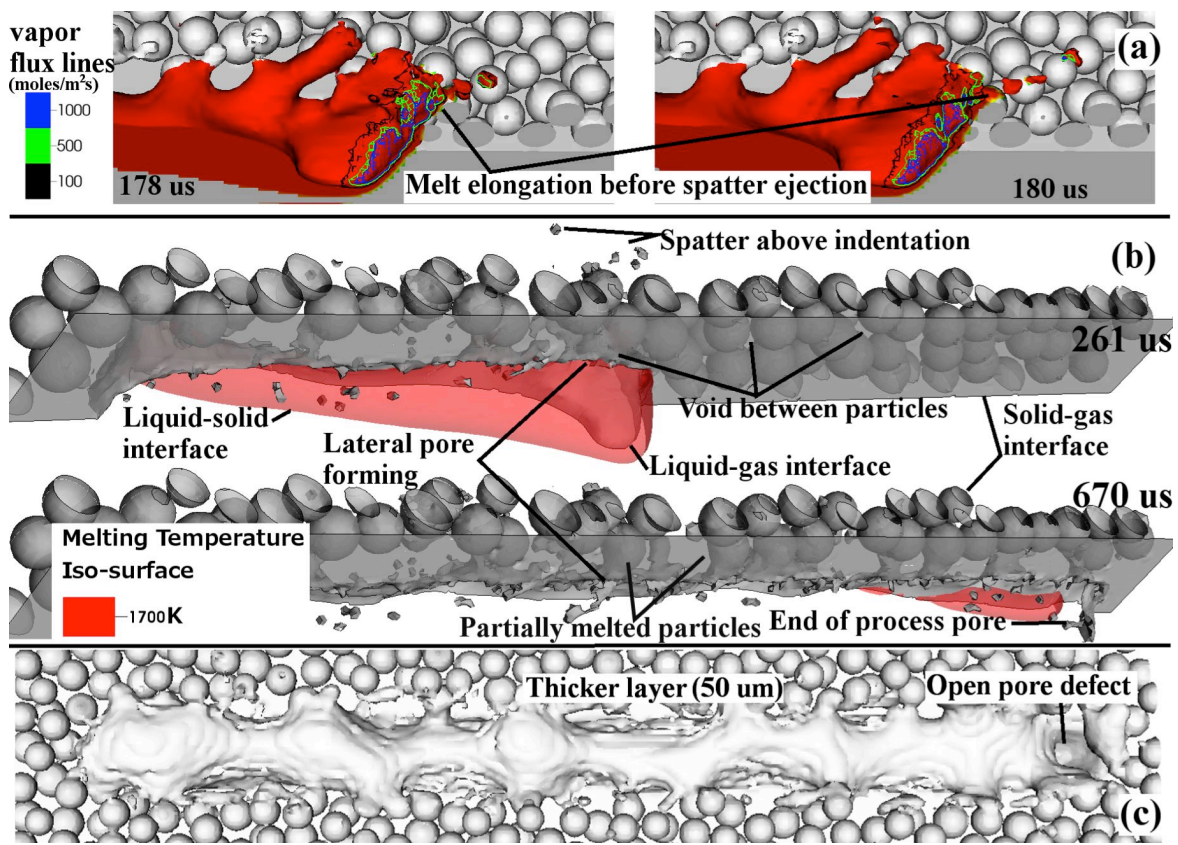

**Figure 6. Formation of defects and spatter.** In a), the 3D selection clips show an elongated fluid column breaking into spatter due to high vapor flux, i.e. evaporations. In b), the snapshots reveal a lateral pore forming out of the voids that exist between the particles. After turning the laser off at 585 μs, an end of process pore forms. It is capped. In c), another end of process pore is shown. But it is open. These pores can seed in more defects in the subsequent layers.

Thijs et al. [19] observed lateral pores when a laser scan was performed with hatch spacing equal to the melt pool width. This means that the neighboring scanning vectors do not overlap each other. They observed the pores between the tracks running parallel to the scan directions. When viewed from the front side of the part, in the direction of the scan, these pores were vertically aligned, and the line repeated in a periodic way along the edges of the melt track width. While these pores are observed at the part level, the defects are seeded at the single layer level [38] and are most likely related to trapped partially melted particles. These pores are certainly undesirable but fortunately, it is possible to eliminate them by appropriately overlapping the neighboring scan tracks. The remedy is to adjust the hatch spacing process parameter to create a 25% scan overlap suggested by Thijs et al. [19].

*3.2.6 End of process pores*

Thijs et al. [20] report on keyhole pores at the end of the scan track. The current model also shows that an opportunity for pores to arise occurs upon switching off the laser. The snapshots taken, after the laser is turned off at 585 μs, in Figures 5 (585-670 μs) and 6b show a large ellipsoidal pore getting trapped beneath the surface due to a fast laser ramp down (1 μs). Two other small spherical pores form this way. Figure 6c offers a different scenario whereby different random powder packing (thicker layer) randomly leaves an uncapped narrow depression.

The remedy for this kind of pores is to allow the surface tension ample time to smooth the surface. So the laser should be ramped down slowly, on the order of few $t_\sigma$ = 27 μs, given by a characteristic time scale for surface tension ($t_\sigma = \sqrt{\rho L^3 / \sigma}$, where $\rho$ is density, $\sigma$ surface tension and L a characteristic length scale).

## 4- Conclusion

In conclusion, this study demonstrates the importance of recoil pressure and Marangoni convection in shaping the melt pool flow and how denudation, spattering, and pore defects emerge and become part of a laser bed-fusion process. The physics processes involved are intimately coupled to each other since they all have a strong dependence on the temperature.

While radiation cooling scales as $T^4$, the evaporative cooling is more efficient at limiting the peak surface temperature because of its exponential dependence on T. This has a strong effect on the magnitude of the recoil pressure since the latter also grows exponentially with the temperature. The recoil force overcomes the surface tension, which opposes the compressive effect of the recoil force, and therefore creates the depression and material spatter. Upon cooling below the boiling point, the surface tension takes over and causes pores to form upon depression wall collapse. The surface tension effects dominate in the transition region where a strong flow (Marangoni effect) takes place. This flow helps with cooling of the depression, creating the denudation zone, pulling in adjacent particles and creating side pores close to partially melted particles. Eventually the transition zone thins out due to the melt flow breaking up and forming the tail-end region. The latter is subject to irregular flow that is short lived due to the drop in temperatures and solidification.

Deep and narrow depressions should be avoided in order to decrease pore formation due to depression collapse. One should also note that, upon changing direction along a scan track, the laser intensity should be decreased otherwise, extra heat deposited could lead to a deep and narrow depression, which collapses and forms pores. An appropriate scan vector overlap can increase the densification by eliminating partially melted and trapped particles and any associated shallow lateral pores. Also, a gentle ramping down of the laser power, on the order of few t$_\sigma$, can prevent end of track pores and side surface roughness.

## 5- Acknowledgment

We acknowledge valuable input from Wayne King. This work was performed under the auspices of the U.S. Department of Energy by Lawrence Livermore National Laboratory under Contract DE-AC52-07NA27344. This work was funded by the Laboratory Directed Research and Development Program under project tracking code 13-SI-002. The LLNL document review and release number is LLNL-JRNL-676495.

## 6- References